\documentclass[preprint]{aastex}

\begin{document}

\title {Absolute dimensions and evolutionary status of the semi-detached Algol W Ursae Minoris}
\author{Jang-Ho Park$^{1,2}$, Kyeongsoo Hong$^{1,3}$, Jae-Rim Koo$^{1,4}$, Jae Woo Lee$^{1,5}$, and Chun-Hwey Kim$^2$}
\affil{$^1$Korea Astronomy and Space Science Institute, Daejon 34055, Korea}
\affil{$^2$Department of Astronomy and Space Science, Chungbuk National University, Cheongju 28644, Korea}
\affil{$^3$Institute for Astrophysics, Chungbuk National University, Cheongju 28644, Korea}
\affil{$^4$Department of Astronomy and Space Science, Chungnam National University, Daejeon 34134, Korea}
\affil{$^5$Astronomy and Space Science Major, Korea University of Science and Technology, Daejeon 34113, Korea}
\email{pooh107162@kasi.re.kr, kyeongsoo76@gmail.com, koojr@cnu.ac.kr, jwlee@kasi.re.kr, kimch@chungbuk.ac.kr}

\begin{abstract}
Double-lined eclipsing binaries allow accurate and direct determination of fundamental parameters such as mass and radius
for each component, and they provide important constraints on the stellar structure and evolution models. In this study,
we aim to determine a unique set of binary parameters for the Algol system W UMi and to examine its evolutionary status.
New high-resolution time-series spectroscopic observations were carried out during 14 nights from April 2008 to March 2011,
and a total of 37 spectra were obtained using the Bohyunsan Optical Echelle Spectrograph. We measured the radial velocities (RVs)
for both components, and the effective temperature of the primary star was found to be $T_{\rm eff,1}$ = 9310 $\pm$ 90 K
by a comparison of the observed spectra and the Kurucz models. The physical parameters of W UMi were derived by an analysis of our RV data
together with the multi-band light curves of Devinney et al. (1970). The individual masses, radii, and luminosities of
both components are $M_1$ = 3.68 $\pm$ 0.10 M$_\odot$ and $M_2$ = 1.47 $\pm$ 0.04 M$_\odot$, $R_1$ = 3.88 $\pm$ 0.03 R$_\odot$ and
$R_2$ = 3.13 $\pm$ 0.03 R$_\odot$, and $L_1$ = 102 $\pm$ 1 L$_\odot$ and $L_2$ = 7.3 $\pm$ 0.1 L$_\odot$, respectively.
A comparison of these parameters with theoretical stellar models showed that the primary component lies in the main-sequence
band, while the less massive secondary is noticeably evolved. The results indicate that the initially more massive star became
the present secondary by losing most of its own mass via mass transfer to the companion (present primary).
\end{abstract}

\section{Introduction}

Algol systems are generally semi-detached interacting binaries that are composed of a hotter and more massive main-sequence
primary star and a Roche-lobe filling giant/subgiant secondary of F-K spectral type (Giuricin et al. 1983; Sarna 1993).
They may be produced from initially detached binaries by tidal interaction and mass transfer between the two components,
which have evolved differently to single stars. The originally more massive star evolves past the terminal-age main sequence
(TAMS) and overflows its limiting lobe, while the gainer becomes the current primary component by accreting the mass transferred
via the Lagrange L$_1$ point. Thus, semi-detached Algols are important for understanding the formation and evolution of
stellar systems, as well as astrophysical phenomena such as mass transfer and accretion between component stars, angular
momentum loss via magnetic stellar winds, and magnetic activity in the cool companion (Erdem \& \"Ozt\"urk 2014; \. Ibano\v{g}lu et al. 2006).
For their study, we need to precisely determine the absolute properties of both components in these systems through a detailed analysis of
both multi-band light curves and double-lined radial velocity (RV) data. If the two kinds of observations are of good quality,
mass and radius measurements can be achieved with accuracies better than about 1 \%, which are used as tests of
stellar evolutionary theory and as distance indicators (Torres et al. 2010).

W UMi (BD+86 244, TYC 4651-61-1, 2MASS J16082718+8611595) is an Algol-type eclipsing binary with an orbital period of 1.7011 d.
It was reported to show photometric variability by Astbury (1913) and also Dyson (1913). The spectroscopic observations of
the binary star were seen as single-lined by Joy \& Dustheimer (1935) and Sahade (1945). The former authors classified
the spectral type of the primary component as A4, and determined its velocity semi-amplitude and systemic velocity to be
$K_1$ = 105.5 km s$^{-1}$ and $\gamma = -7.7$ km s$^{-1}$, respectively. On the contrary, the latter author obtained different
velocities of $K_1$ = 86.6 km s$^{-1}$ and $\gamma = -17.7$ km s$^{-1}$. He presented that the primary component is an A3 star
with a rotational velocity of $\sim$40 km s$^{-1}$ and a small eccentricity of $e$ = 0.09. However, Lucy \& Sweeney (1971)
preferred to adopt a circular orbit in their rediscussion of the Sahade's data. From the Hipparcos, ASAS, and NSVS data,
Kreiner et al. (2008) reported that the binary system is in a circular orbit, and its eccentricity is probably spurious.

On the other hand, Devinney et al. (1970) reviewed the observational history of the stellar system prior to 1970 and
separately analyzed two sets of light curves observed in $BV$ filters at the Flower and Cook Observatory (hereafter, $B_{\rm FC}$,
$V_{\rm FC}$) and in $UBV$ filters at the Dyer Observatory (hereafter, $U_{\rm D}$, $B_{\rm D}$, $V_{\rm D}$). They found that
the photometric parameters ($i \approx $ 84 deg, $r_2/r_1 \approx $ 0.78) for the two data sets agreed very well with each other.
The spectral type of the primary component was classified as A3 $\pm$ 2 from the color indices of ($U-B$) and ($B-V$), which is
in good agreement with the spectroscopic classifications. Since then, Mardirossian et al. (1980) and Djurasevic et al. (2003)
re-analyzed the photometric data of Devinney et al. (1970), and both of them presented a mass ratio of $q_{\rm ph} \simeq$ 0.48.
Although the same data were used in the light-curve synthesis, Devinney et al. (1970) and Mardirossian et al. (1980) reported
that W UMi is in detached or semi-detached configurations, while Djurasevic et al. (2003) showed that the eclipsing binary is
a semi-detached system with the secondary component filling its inner Roche lobe.

As previously mentioned, although several photometric studies of W UMi were made by analyzing the multi-band light curves of
Devinney et al. (1970), there are no radial velocities (RVs) for the secondary component. Hence, the spectroscopic mass ratio
and the reliable absolute dimensions have not been established so far. The main purpose of this study is to present the precise
physical properties of the Algol system and to examine its evolutionary state based on our new time-series spectroscopy and
on the existing photometric data. The remainder of this paper is organized as follows. In Section 2, we describe new spectroscopic observations
and data analysis of our program target. The absolute dimensions of each component are determined from the binary modeling
in Section 3. In Section 4, the results are summarized and the evolutionary state is discussed.

\section{Observations and Data Analysis}

New spectroscopic observations were made on 14 nights between April 2008 and March 2011 using the 1.8 m telescope at
the Bohyunsan Optical Astronomy Observatory (BOAO). We obtained a total of 37 spectra using the Bohyunsan Optical Echelle
Spectrograph (BOES) attached to the telescope, which is designed to cover the wavelength region of 3600$-$10200 \AA.
More information about the specifications of the BOES can be found in the paper of Kim et al. (2007). We used a resolving power
($\lambda/\Delta\lambda$) of 30000 with the largest fiber of 300 $\mu$m diameter.  The observed spectra were acquired with
an exposure time of 2700 s, and they were pre-processed with the IRAF/CCDRED package and extracted to one-dimensional spectra with
the IRAF/ECHELLE package. The typical signal-to-noise (S/N) ratio at $5000-6000$ {\AA} was approximately 40.

To measure the RVs for the primary and secondary stars in the observed spectra, we searched for absorption lines that
were strong enough that both components could be clearly identified. Although the temperature difference between them is
large ($\Delta T$ $\approx$ 3800 K) and the luminosity contribution of the secondary star to the W UMi system is very low
(Djurasevic et al. 2003), we found that the isolated line of Fe I 4957.61 {\AA} in our spectra shows the spectral line from the faint secondary.
Some series spectra in the Iron region are shown in Figure 1. The RVs of each component were determined using the line profile fitting with
two Gaussian functions (Hong et al. 2015), and the resultant RVs and their errors are listed in Table 1.

The effective temperature ($T_{\rm eff,1}$) and the projected rotational velocity ($v_1 \sin i$) of the primary star were
determined by the $\chi^2$ fitting method (Guo et al. 2016; Hong et al. 2017), which is used to minimize the difference between
observed spectra and theoretical stellar atmosphere models. As mentioned in the Introduction, the primary component of W UMi
was classified as an A-type star from its single-lined spectra and color indices. Thus, we selected the Fe II $\lambda$ 4046,
Fe I $\lambda$ 4957, Ca I $\lambda$ 4427, and Mg II $\lambda$ 4481 regions proper for temperature indicators of A-type dwarfs
according to the {\it Digital Spectral Classification Atlas} of R. O. Gray.\footnote{More information is available on the
website (https://ned.ipac.caltech.edu/level5/Gray/frames.html).}
First of all, we used the spectral disentangling code FDBinary (Iliji\'c et al. 2004) to obtain the reconstructed spectra of
the four regions from all observed spectra. Then, 5850 synthetic spectra with ranges of
${T_{\rm eff}= 8760-10250}$ K and $v_1 \sin i = 96-134$ km s$^{-1}$ were interpolated from the grids of atmosphere models by Kurucz (1993).
For this process, the solar metallicity of $[$Fe/H$]=0$ and a microturbulent velocity of 2.0 km s$^{-1}$ were assumed.
The surface gravity of the primary component was set to be $\log$ $g_1$ = 3.8 from our binary solutions, which is obtained in the following section.
Finally, we calculated the $\chi^2$ values between the reconstructed and synthetic spectra in these regions, and the search results are shown in Figure 2.
The best-fitting parameters were determined to be $T_{\rm eff,1}$ = 9310 $\pm$ 90 K and $v_1 \sin i_{1}$ = 105 $\pm$ 6 km s$^{-1}$,
respectively, and their errors were calculated from the standard deviations of the best values obtained in each region.
The observed spectra of Fe II $\lambda$ 4046, Fe I $\lambda$ 4957, Ca I $\lambda$ 4427, and Mg II $\lambda$ 4481
are presented in Figure 3, together with three synthetic spectra of 9400 K, 9310 K, and 9220 K, respectively.

Because the secondary star contributes only a few percent to the total luminosity of the W UMi system, and the S/N ratios of
the observed spectra were not enough, it is difficult to reconstruct the spectrum of the cool component and to directly obtain
its atmospheric parameters. Therefore, we checked the atmospheric parameters of the secondary star by comparing the observed
spectra with the synthetic spectra computed from the temperature of $T_{\rm eff,2}= 5370$ K and the surface gravity of
$\log$ $g_2$ = 3.6 presented in Section 3.
The results are shown in Figure 1, where the blue and red lines
indicate the synthetic spectra of the primary and secondary stars, respectively. As seen in the figure,
the observed spectra are in satisfactory agreement with the synthetic spectra of both components.

\section{Binary Modeling and Absolute Dimensions}

To obtain the binary parameters of W UMi, our double-lined RV curves were analyzed together with the $(BV)_{\rm FC}(UBV)_{\rm D}$
light curves of Devinney et al. (1970) using the 2003 version of the Wilson-Devinney synthesis code (Wilson \& Devinney 1971;
van Hamme \& Wilson 2003; hereafter W-D). In this paper, we refer to the primary and secondary components as those being
eclipsed at Min I (phase 0.0) and Min II, respectively. The effective temperature of the hotter and more massive primary star
was fixed at $T_{\rm 1}$ = 9310 K from our spectral analyses. We adopted the gravity-darkening exponents and bolometric albedoes
for each component to be ($g_{1}$, $A_{1}$) = (1.0, 1.0) and ($g_{2}$, $A_{2}$) = (0.32, 0.5), which are appropriate for stars
with radiative and convective envelopes, respectively. Linear bolometric ($X$, $Y$) and monochromatic ($x$, $y$) limb-darkening
coefficients were interpolated from the values of van Hamme (1993) in concert with the model atmosphere option.
Recently, Kreiner et al. (2008) proposed the existence of a circumbinary object in the system from an eclipse timing analysis.
Thus, we looked for a possible third light ($\ell_{3}$), but we found that the parameter remains zero within its error. We set
the third light to be $\ell_{3}$ = 0.0 for further analyses.

The time gap between the photometric and spectroscopic data used in our synthesis is very long, and the orbital period has
varied due to a sinusoidal variation superposed on a downward parabola (Kreiner et al. 2008). Thus, the analyses of these data
sets were separately performed in two steps, following the procedure described by Hong et al. (2015) and Koo et al. (2016).
In the first step, our double-lined RV curves were modeled by adjusting the epoch ($T_0$), the orbital period ($P$),
the semi-major axis ($a$), the systemic velocity ($\gamma$), and the mass ratio ($q$), where the orbital ephemeris were
initialized with the linear elements calculated from the eclipse timings of W UMi. In the second step, we analyzed the multi-band
light curves of Devinney et al. (1970) with the spectroscopic elements obtained in the first stage. These steps were repeated
until the binary parameters of W UMi were unchanged. Our binary modeling started from the detached mode 2, but converged to
the semi-detached mode 5 (the secondary filling its inner Roche lobe) during the computation. The final results are listed in
Table 2, where $r$ (volume) is the mean volume radius calculated from the tables of Mochnacki (1984). Figure 4 shows
the RV curves of W UMi with the model fits, wherein the measurements from Joy \& Dustheimer (1935) and Sahade (1945) are plotted
together for comparison. Figure 5 displays the $(BV)_{\rm FC}(UBV)_{\rm D}$ light curves of Devinney et al. (1970) with the W-D fits.

Our RV and light solution indicates that W UMi is a semi-detached Algol with parameters of $q$ = 0.399, $i$ = 86$^\circ$.0,
and $\Delta$($T_{1}$--$T_{2}$) = 3944 K, where the primary star fills its Roche lobe by 77\%. From the consistent set of
binary parameters, the absolute dimensions for each component were calculated using the JKTABSDIM code (Southworth et al. 2005),
and they are listed in Table 3. The luminosity ($L$) and bolometric magnitudes ($M_{\rm bol}$) were computed by the adoption of
$T_{\rm eff}$$_\odot$ = 5780 K and $M_{\rm bol}$$_\odot$ = +4.77 for solar values. For the absolute visual magnitudes ($M_{V}$),
we used the bolometric corrections (BCs) from Girardi et al. (2002) appropriate for the effective temperatures of each component.
The mass ($M_1$ = 3.68 M$_\odot$) of the primary star is much heavier than that ($\sim$ 1.5 M$_\odot$) of normal main-sequence
stars with the same effective temperature, but is compatible with the mass-luminosity relation of $L_{1}$ $\propto$ $M^{3.20}$
for the semi-detached binaries given by \. Ibano{\v g}lu et al. (2006). This may be because W UMi has experienced a different
evolutionary process than single stars through the secondary to primary mass transfer.

Using an apparent visual magnitude of $V$ = 8.61 at maximum light (Kreiner et al. 2008) and an interstellar absorption of
$A_{V}$ = 0.382 from the Galactic 3D model (Drimmel et al. 2003), we computed the distance of the W UMi system to be
488 $\pm$ 10 pc. This result is in excellent agreement with the value of 506 $\pm$ 63 pc obtained by the trigonometric parallax
(1.98 $\pm$ 0.25 mas) from {\it Gaia} DR1 (Gaia Collaboration et al. 2016).

\section{Summary and Discussion}

In this article, high-resolution time-series spectroscopic observations were presented for the Algol system W UMi. We analyzed
in detail the new spectra with existing photometric data. The results from this work can be summarized as follows:

\begin{enumerate}

\item From the observed spectra, we detected an isolated absorption line (Fe I 4957.61 {\AA}) of the cool secondary star
and measured the RVs of both components; the secondary's features are the first ever found. The spectroscopic mass ratio of
the eclipsing binary was calculated to be $q = 0.399 \pm 0.009$ from the RV semi-amplitudes $K_1$ and $K_2$, which is
remarkably smaller than the photometric results of $q_{\rm ph} \simeq$ 0.48 (Mardirossian et al. 1980; Djurasevic et al. 2003).
Because it was obtained from the double-lined RV curves, our $q$ value should be more accurate and reliable than that
obtained from only the photometric data.

\item The effective temperature and the projected rotational velocity of the primary star were determined to be $T_{\rm eff,1}$
= 9310 $\pm$ 90 K and $v_1 \sin i_{1}$ = 105 $\pm$ 6 km s$^{-1}$, respectively, by minimizing the $\chi^2$ values between
the reconstructed and synthetic spectra in four regions. The measured velocity indicates that the primary star may be in
a synchronous rotation ($v_{\rm 1,sync}$ = 115.3 $\pm$ 1.0 km s$^{-1}$) with the orbital motion. On the other hand, it is
difficult to obtain the atmospheric parameters of the secondary star, because the S/N ratios of the observed spectra and
the light ratio of the components are low. Thus, we compared the observed spectra with the synthetic spectra from
the secondary's parameters ($T_{\rm eff,2}= 5370$ K and $\log$ $g_2$ = 3.6) yielded in this study. We can see that there is
a satisfactory agreement between them.

\item Our double-lined RVs were solved with the multi-band light curves given by Devinnet et al. (1970). The analysis results
demonstrate that W UMi is a classical Algol system in which the primary component fills its limiting Roche lobe by 77\% and
is slightly larger than the lobe-filling secondary. The absolute dimensions of each component were calculated from the RV
and light parameters. The masses of the primary and secondary stars were determined to be $M_1$ = 3.68 $\pm$ 0.10 M$_\odot$
and $M_2$ = 1.47 $\pm$ 0.04 M$_\odot$, and their radii are $R_1$ = 3.88 $\pm$ 0.03 R$_\odot$ and $R_2$ = 3.13 $\pm$ 0.03 R$_\odot$.
The uncertainties of the masses and radii are about 3\% and 1\%, respectively.

\end{enumerate}

The locations of both components of W UMi are shown in the Hertzsprung-Russell (HR) diagram of Figure 6, together with those
of 60 other semi-detached Algols (\. Ibano\v{g}lu et al. 2006). Here, the primary component lies in the main-sequence band,
while the secondary component is beyond TAMS, a location where the secondaries of many Algol systems are found. To examine
the evolutionary state of W UMi, we display a conservative evolutionary track for a binary star, consisting of a mass-donor component
(low-mass progenitor) of 3.0 M$_\odot$ and its companion (present primary) of 1.2 M$_\odot$ with an initial orbital period
of 2.0 d (De Loore \& van Rensbergen 2005). As shown in Figure 6, the primary (P) and
secondary (S) components of W UMi are a good match to the conservative binary evolution model.
Further, they lie on the same isochrone with the present age of about 0.36 Gyr.
Although the binary evolution tracks are not for a stellar system with the same total mass of 5.15 M$_\odot$ as our program target,
it is thought that W UMi has undergone conservative binary evolution through a case A mass transfer between the component stars.
On this account, the present primary star was formed by mass accretion from the initially more massive component, and
the donor star became the evolved low-mass secondary star by transferring most of its own mass to the gainer (present primary).

\acknowledgments{ }
This research was supported by Korea Astronomy and Space Science Institute (KASI) grant 2018-1-830-02. The work by K. Hong and
C.-H. Kim, and J.-R. Koo was supported by the grant numbers 2017R1A4A1015178 and 2017R1A6A3A01002871 of the National Research Foundation (NRF)
of Korea, respectively. We have used the Simbad database maintained at CDS, Strasbourg, France, and data from the European Space Agency (ESA)
mission {\it Gaia} (\url{https://www.cosmos.esa.int/gaia}), processed by the {\it Gaia} Data Processing and Analysis Consortium\newline
(DPAC,\url{https://www.cosmos.esa.int/web/gaia/dpac/consortium}). Funding for the DPAC has been provided by national institutions,
in particular the institutions participating in the {\it Gaia} Multilateral Agreement.

\clearpage
\begin{figure}
\centering
\includegraphics[]{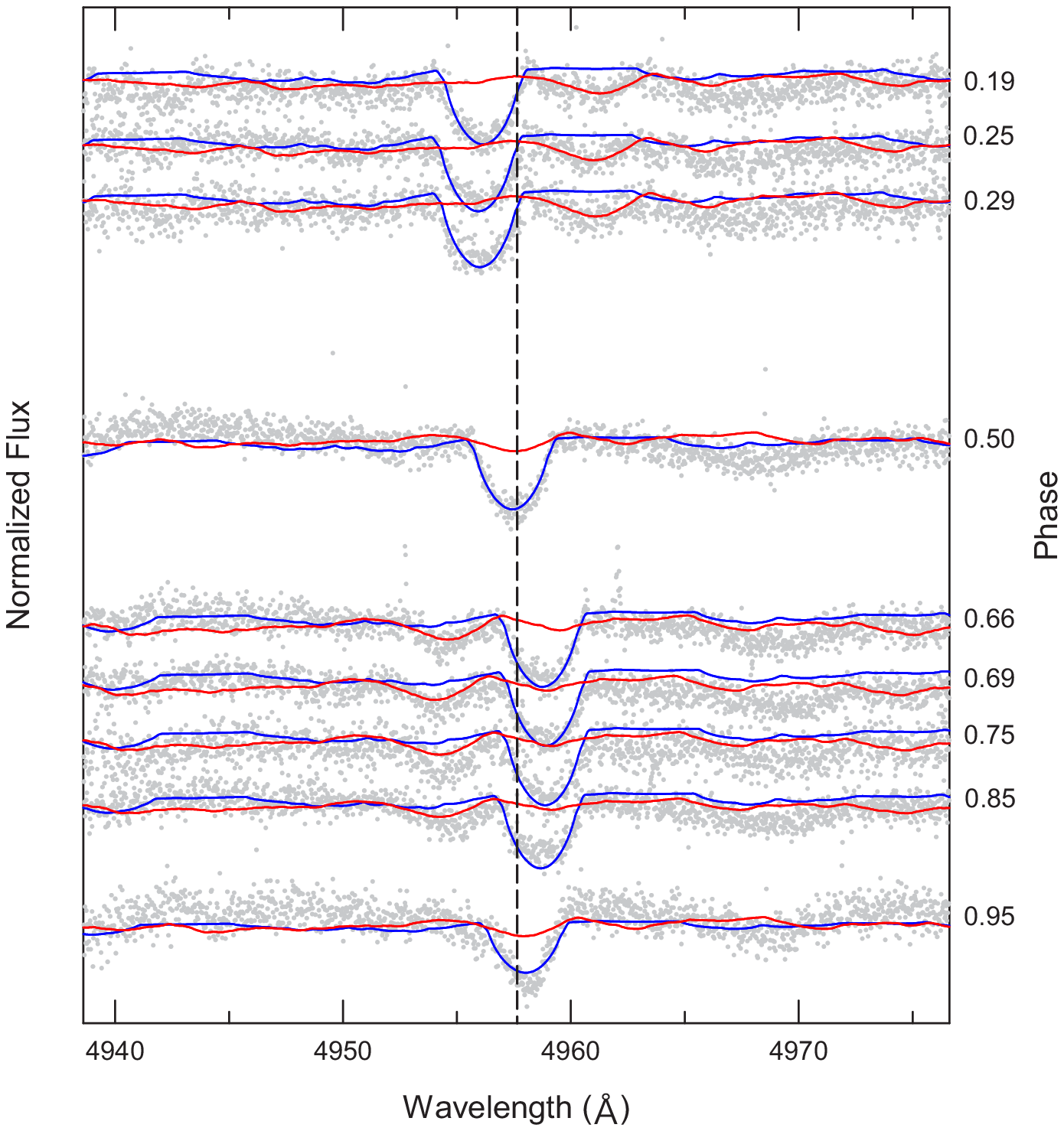}
\caption{Series spectral profiles of the Fe I $\lambda$ 4957.61 region for W UMi. The orbital phase is indicated for
each spectrum. A vertical line indicates the central wavelength of the Fe I line.
The circles represent the observed spectra, and the blue and red lines represent
the synthetic spectra of the primary ($T_{\rm eff,1}$ = 9310 K, $\log$ $g_1$ = 3.8) and
secondary components ($T_{\rm eff,2}$ = 5370 K, $\log$ $g_2$ = 3.6), respectively.}
\label{Fig1}
\end{figure}

\begin{figure}
\centering
\includegraphics[width=1\columnwidth]{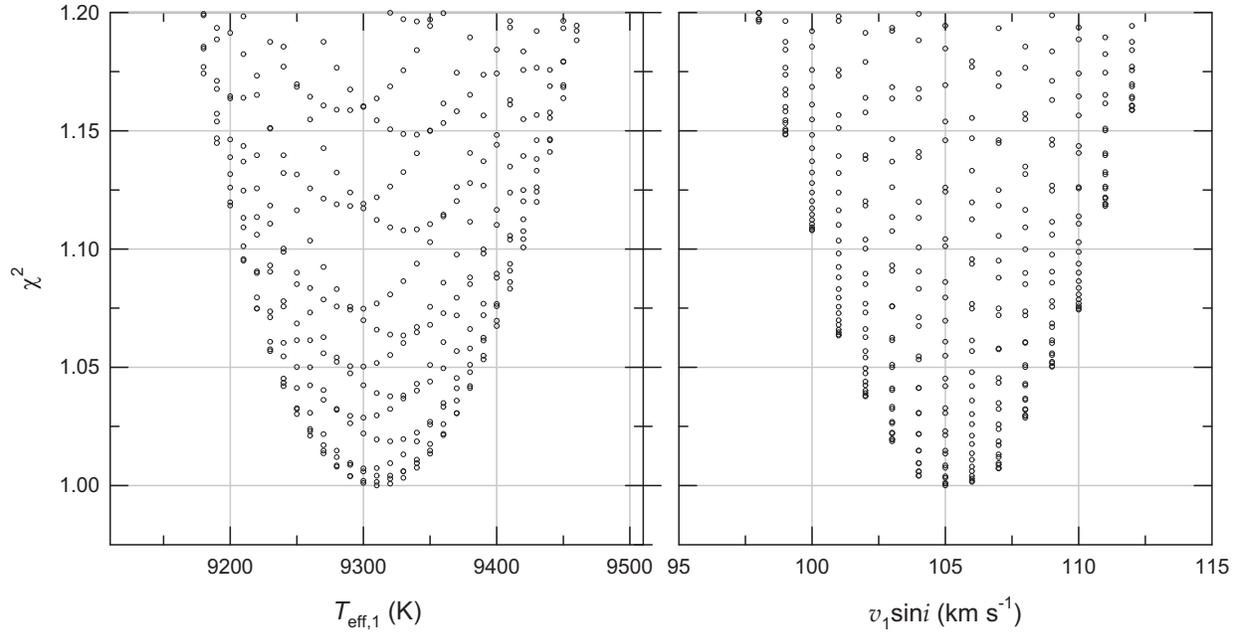}
\caption{$\chi^2$ diagrams of the effective temperature ($T_{\rm eff,1}$) and the projected rotational velocity ($v_1 \sin i$)
of the primary star.}
\label{Fig2}
\end{figure}

\begin{figure}
\centering
\includegraphics[]{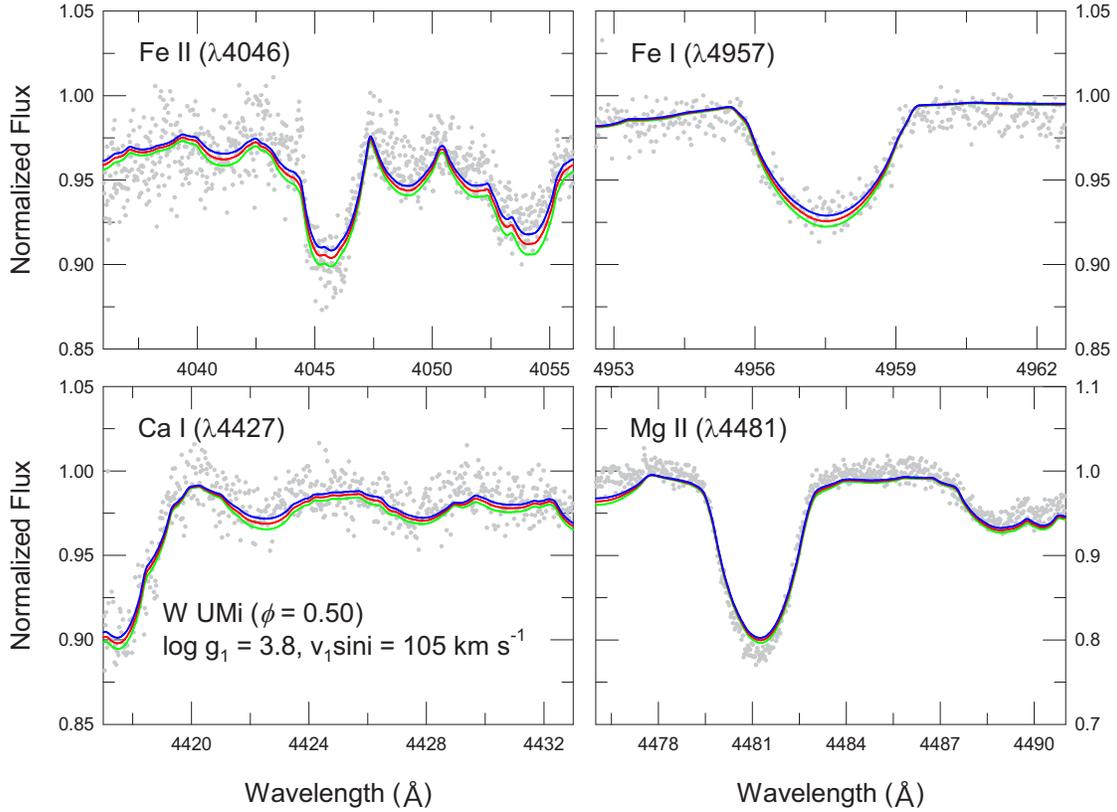}
\caption{Four spectral regions of the primary star. The circles are the spectrum observed at an orbital phase of $\phi$ = 0.50.
The blue, red, and green lines represent the synthetic spectra of 9400 K, 9310 K, and 9220 K, respectively,
from the atmosphere models of Kurucz (1993).}
\label{Fig3}
\end{figure}

\begin{figure}
\centering
\includegraphics[]{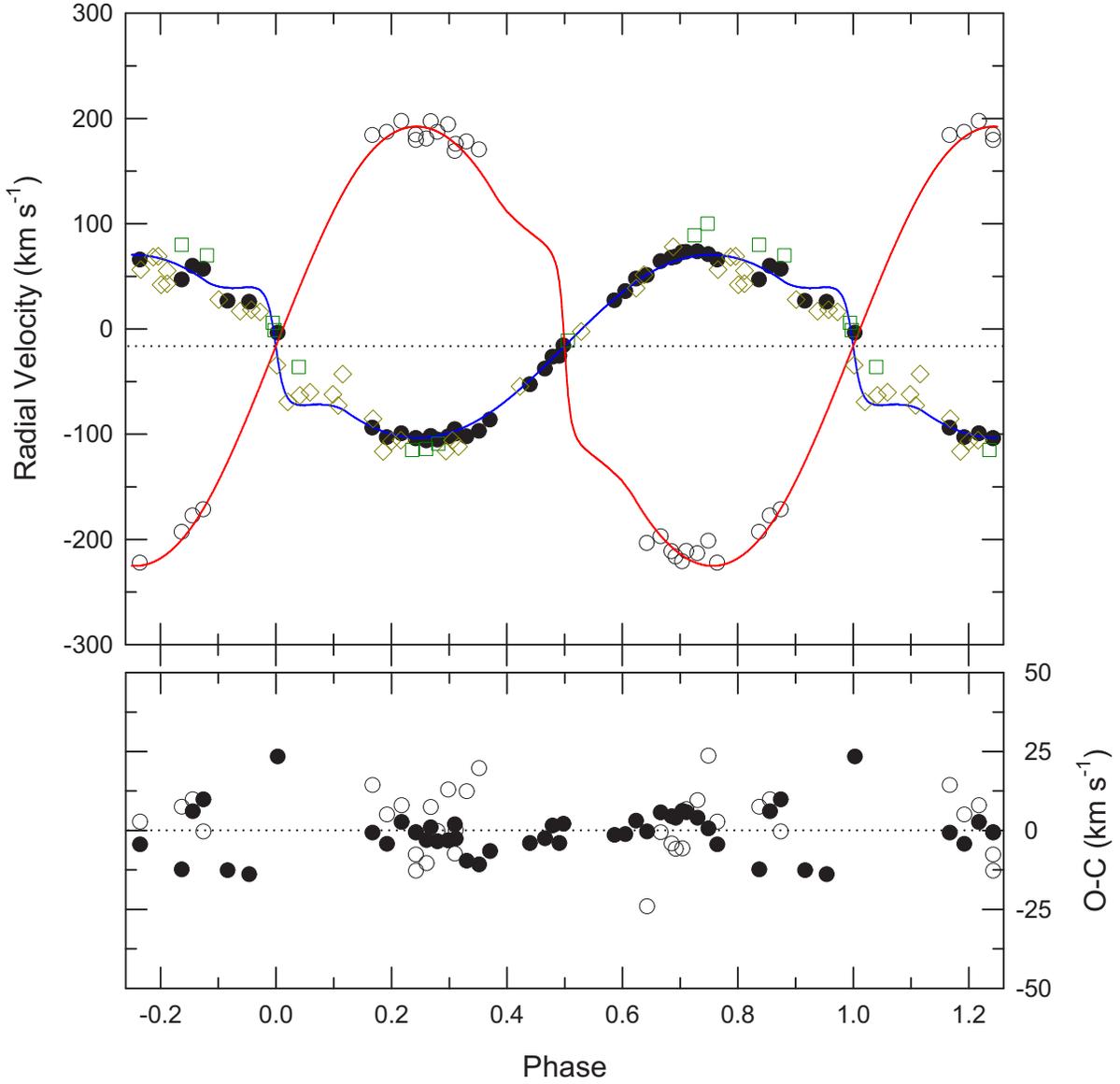}
\caption{RV curves of W UMi with fitted models. The filled and open circles represent our double-lined RV measurements for
the primary and secondary components, respectively. The square and diamond symbols represent the single-lined RVs of
Joy \& Dustheimer (1935) and Sahade (1945), respectively. In the upper panel, the solid curves denote the results
from a consistent light and RV curve analysis with the W-D code. The dotted line represents the system velocity of
$-$15.4 km s$^{-1}$. The lower panel shows the residuals between observations and theoretical models.}
\label{Fig4}
\end{figure}

\begin{figure}
\centering
\includegraphics[]{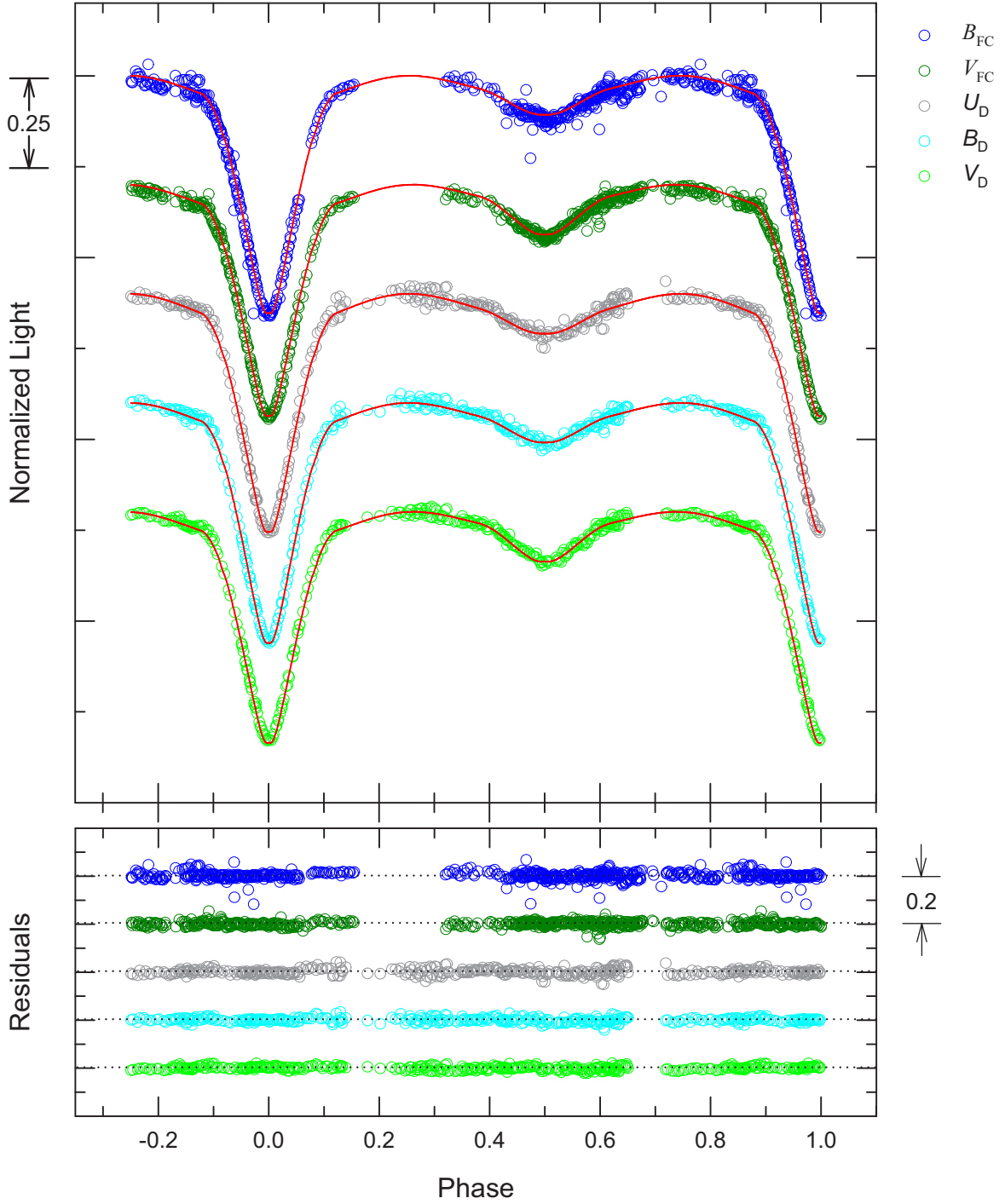}
\caption{$(BV)_{\rm FC}(UBV)_{\rm D}$ light curves of W UMi with fitted models. The circles are individual measures taken
from Devinney et al. (1970), and the solid lines represent the synthetic curves obtained with the W-D runs. The lower panel
shows the light residuals between measurements and theoretical models.}
\label{Fig5}
\end{figure}

\begin{figure}
\centering
\includegraphics[]{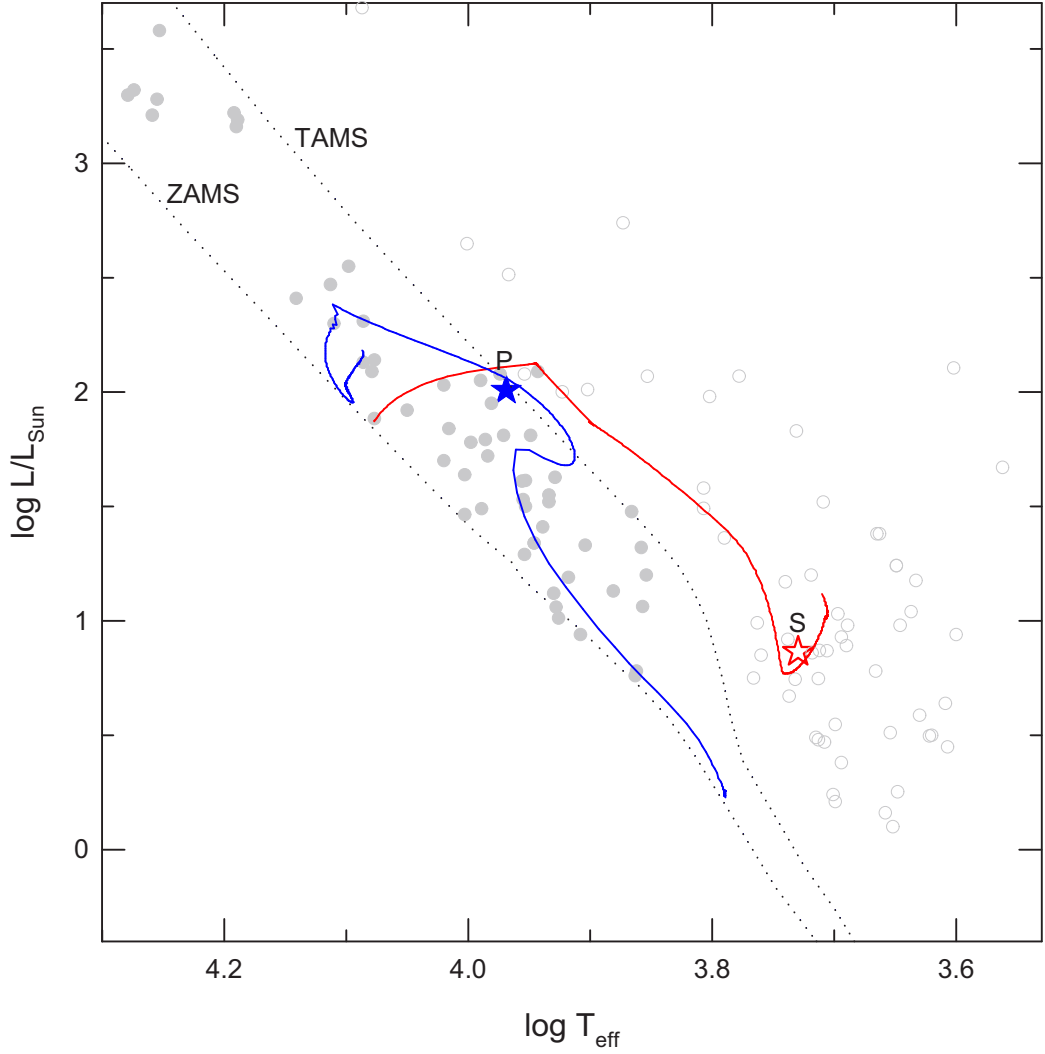}
\caption{Positions on the HR diagram for W UMi (star symbols) and other semi-detached Algols (circles, \. Ibano{\v g}lu et al. 2006).
The filled and open symbols represent the primary and secondary stars, respectively. The red and blue lines
denote the conservative evolutionary track for a binary with a total mass of 4.2 M$_\odot$ (De Loore \& van Rensbergen 2005).}
\label{Fig6}
\end{figure}

\clearpage
\begin{deluxetable}{lcrrrr}
\tabletypesize{\scriptsize}
\tablewidth{0pt}
\tablecaption{Radial Velocities of W UMi.}
\tablehead{
\colhead{HJD}          & \colhead{Phase}        & \colhead{$V_{1}$}      & \colhead{$\sigma_{1}$} & \colhead{$V_{2}$}      & \colhead{$\sigma_{2}$} \\
\colhead{(2450000+)}   &                        & \colhead{km s$^{-1}$}  & \colhead{km s$^{-1}$}  & \colhead{km s$^{-1}$}  & \colhead{km s$^{-1}$}
}
\startdata
4568.0616              &  0.440                 &  $-$52.6               &      2.5               &  \dots                 &  \dots                 \\
4568.1053              &  0.466                 &  $-$37.7               &      3.1               &  \dots                 &  \dots                 \\
4568.1481              &  0.491                 &  $-$25.5               &      4.6               &  \dots                 &  \dots                 \\
4570.9995              &  0.167                 &  $-$93.9               &      1.3               &    184.2               &      1.6               \\
4571.0423              &  0.192                 & $-$102.8               &      3.4               &    187.3               &      5.8               \\
4571.0852              &  0.217                 &  $-$99.3               &      1.7               &    197.7               &      1.4               \\
4571.1277              &  0.242                 & $-$103.9               &      1.2               &    179.5               &      8.4               \\
4571.1714              &  0.268                 & $-$101.6               &      0.8               &    197.5               &      3.9               \\
4572.0157              &  0.764                 &     65.9               &      1.3               & $-$222.2               &      8.8               \\
5281.2352              &  0.692                 &     68.7               &      1.3               & $-$216.2               &      1.4               \\
5281.2670              &  0.711                 &     73.4               &      3.0               & $-$211.0               &      2.7               \\
5281.2996              &  0.730                 &     73.6               &      1.1               & $-$213.1               &      1.8               \\
5281.3313              &  0.749                 &     71.0               &      2.0               & $-$201.3               &      2.4               \\
5282.2880              &  0.311                 &  $-$99.4               &      1.2               &    175.9               &      3.3               \\
5282.3198              &  0.330                 & $-$102.1               &      5.1               &    178.1               &      2.4               \\
5284.2744              &  0.479                 &  $-$26.3               &      1.2               &  \dots                 &  \dots                 \\
5284.3062              &  0.498                 &  $-$15.6               &      1.1               &  \dots                 &  \dots                 \\
5578.3414              &  0.352                 &  $-$96.9               &      1.9               &    170.5               &      0.9               \\
5578.3733              &  0.371                 &  $-$86.2               &      2.8               &  \dots                 &  \dots                 \\
5579.3014              &  0.916                 &     26.7               &      1.2               &  \dots                 &  \dots                 \\
5579.3652              &  0.954                 &     25.8               &      1.3               &  \dots                 &  \dots                 \\
5628.2076              &  0.667                 &     64.4               &      1.8               & $-$197.0               &      3.2               \\
5628.2393              &  0.685                 &     67.7               &      2.3               & $-$211.1               &      4.2               \\
5628.2710              &  0.704                 &     72.8               &      2.1               & $-$220.9               &      2.2               \\
5629.1863              &  0.242                 & $-$103.9               &      3.5               &    184.7               &      5.2               \\
5629.2180              &  0.261                 & $-$106.1               &      2.1               &    180.8               &      3.6               \\
5629.2499              &  0.279                 & $-$105.1               &      2.6               &    187.4               &      6.8               \\
5629.2816              &  0.298                 & $-$102.3               &      4.2               &    194.5               &      1.9               \\
5630.1983              &  0.837                 &     46.9               &      3.8               & $-$192.8               &      7.0               \\
5630.2300              &  0.856                 &     59.9               &      2.4               & $-$177.2               &      4.7               \\
5630.2619              &  0.874                 &     57.1               &      0.9               & $-$171.4               &      3.6               \\
5631.0028              &  0.310                 &  $-$95.1               &      1.6               &    169.1               &      6.5               \\
5632.1817              &  0.003                 &   $-$3.6               &      4.2               &  \dots                 &  \dots                 \\
5633.1746              &  0.587                 &     27.3               &      1.1               &  \dots                 &  \dots                 \\
5633.2063              &  0.605                 &     35.9               &      1.1               &  \dots                 &  \dots                 \\
5633.2380              &  0.624                 &     47.8               &      2.8               &  \dots                 &  \dots                 \\
5633.2699              &  0.643                 &     51.1               &      2.1               & $-$203.5               &      4.1               \\
\enddata
\end{deluxetable}

\clearpage
\begin{deluxetable}{lcc}
\tablewidth{0pt}
\tablecaption{Light and RV Parameters of W UMi.}
\tablehead{
\colhead{Parameter}                      & \colhead{Primary}   & \colhead{Secondary}
}
\startdata
$T_0$ (HJD)                              & \multicolumn{2}{c}{2439758.8456 $\pm$ 0.0001}     \\
$P$ (d)                                  & \multicolumn{2}{c}{1.7011577 $\pm$ 0.0000001}     \\
$i$ (deg)                                & \multicolumn{2}{c}{86.0 $\pm$ 0.1}                \\
$T$ (K)                                  & 9310 $\pm$ 90       & 5366 $\pm$ 17               \\
$\Omega$                                 & 3.139 $\pm$ 0.003   & 2.677                       \\
$\Omega_{\rm in}$                        & \multicolumn{2}{c}{2.677}                         \\
$l$/($l_{1}$+$l_{2}$){$_{B_{\rm FC}}$}   & 0.9675 $\pm$ 0.0009 & 0.0325                      \\
$l$/($l_{1}$+$l_{2}$){$_{V_{\rm FC}}$}   & 0.9340 $\pm$ 0.0005 & 0.0660                      \\
$l$/($l_{1}$+$l_{2}$){$_{U_{\rm D}}$}    & 0.9764 $\pm$ 0.0009 & 0.0236                      \\
$l$/($l_{1}$+$l_{2}$){$_{B_{\rm D}}$}    & 0.9675 $\pm$ 0.0007 & 0.0325                      \\
$l$/($l_{1}$+$l_{2}$){$_{V_{\rm D}}$}    & 0.9340 $\pm$ 0.0005 & 0.0660                      \\
$r$ (pole)                               & 0.3618 $\pm$ 0.0004 & 0.2824 $\pm$ 0.0003         \\
$r$ (point)                              & 0.3958 $\pm$ 0.0007 & 0.4069 $\pm$ 0.0015         \\
$r$ (side)                               & 0.3750 $\pm$ 0.0005 & 0.2944 $\pm$ 0.0004         \\
$r$ (back)                               & 0.3857 $\pm$ 0.0006 & 0.3270 $\pm$ 0.0004         \\
$r$ (volume)                             & 0.3745 $\pm$ 0.0005 & 0.3025 $\pm$ 0.0004         \\
\multicolumn{3}{l}{Spectroscopic orbits:}                                                    \\
$T_0$ (HJD)                              & \multicolumn{2}{c}{2454385.300 $\pm$ 0.013}       \\
$P$ (d)                                  & \multicolumn{2}{c}{1.701060 $\pm$ 0.000019}       \\
$a$ (R$_\odot$)                          & \multicolumn{2}{c}{10.355 $\pm$ 0.093}            \\
$\gamma$ (km s$^{-1}$)                   & \multicolumn{2}{c}{$-$16.4 $\pm$ 1.1}             \\
$K_{1}$ (km s$^{-1}$)                    & \multicolumn{2}{c}{87.6 $\pm$ 1.3}                \\
$K_{2}$ (km s$^{-1}$)                    & \multicolumn{2}{c}{219.5 $\pm$ 2.3}               \\
$q$                                      & \multicolumn{2}{c}{0.399 $\pm$ 0.009}             \\
\enddata
\end{deluxetable}

\clearpage
\begin{deluxetable}{lccc}
\tablewidth{0pt}
\tablecaption{Physical Properties of W UMi.}
\tablehead{
\colhead{Parameter}       & \colhead{Primary}     & \colhead{Secondary}
}
\startdata
$M$ (M$_\odot$)           & 3.68  $\pm$ 0.10     & 1.47  $\pm$ 0.04       \\
$R$ (R$_\odot$)           & 3.88  $\pm$ 0.03     & 3.13  $\pm$ 0.03       \\
$\log$ $L/L_\odot$        & 2.01  $\pm$ 0.02     & 0.87  $\pm$ 0.01       \\
$\log$ $g$ (cgs)          & 3.83  $\pm$ 0.01     & 3.61  $\pm$ 0.01       \\
$M_{\rm bol}$ (mag)       & $-$0.27 $\pm$ 0.05   & $+$2.59 $\pm$ 0.02     \\
BC (mag)                  & $-$0.12              & $-$0.14                \\
$M_{\rm V}$ (mag)         & $-$0.15 $\pm$ 0.03   & $+$2.73 $\pm$ 0.03     \\
Distance (pc)             & \multicolumn{2}{c}{488 $\pm$ 10}              \\
\enddata
\end{deluxetable}

\end{document}